\documentclass[twocolumn,showpacs,preprintnumbers,amsmath,amssymb,floatfix]{revtex4}
\usepackage{graphicx}
\usepackage{dcolumn}
\usepackage{bm}
\usepackage{citesort} 
\usepackage{array,rotating}
\usepackage[dvips]{color}
\usepackage{relsize}

\begin{document}


\title{Thermodynamic Properties of Spin Ladders with Cyclic Exchange}

\author{Alexander B\"uhler} 
 \email{ab@thp.uni-koeln.de} 
\author{Ute L\"ow} 
 \email{ul@thp.uni-koeln.de} 
\author{Kai P. Schmidt} 
 \email{ks@thp.uni-koeln.de} 
\author{G\"otz S. Uhrig}
 \homepage{http://www.thp.uni-koeln.de/~gu}
 \email{gu@thp.uni-koeln.de}
\affiliation{Institut f\"ur Theoretische Physik, %
   Universit\"at zu K\"oln, %
   Z\"ulpicher Stra{\ss}e 77, %
   50937 K\"oln, %
   Germany}

\date{\today}

\begin{abstract}
By high temperature series expansion and exact complete diagonalization the
magnetic susceptibility $\chi(T)$ and the specific heat $C(T)$ of a
two-leg $S=1/2$ ladder with cyclic (4-spin) exchange are
computed. Both methods yield convincing results for not too small
temperatures. We find that a small amount of cyclic exchange influences the
thermodynamical properties significantly. Our results can serve as
reliable basis for an efficient analysis of experimental data
\end{abstract}

\pacs{05.10.-a, 75.40.Cx, 75.10.Jm, 75.50.Ee}

\maketitle

\section{Introduction\label{sec:intro}}
Recently, it has been pointed out that both for spin ladder systems
and the parent compounds of high-T$_{c}$ superconductors
\cite{dagot96} in addition to the bilinear exchange also biquadratic
ring exchange terms are important
\cite{schmi90,honda93,brehm99,reisc01,senth01a,balen02}. In Refs.
\cite{loren99,mizun99,matsu00b,limin00,schmi01,katan01,colde01,windt01,nunne02}
the modification of the low temperature behavior due to this new
exchange interaction has been discussed in detail.  However, only
little investigations \cite{johns00a} of the impact of this new type
of interaction on the finite temperature properties are available up
to now.  Therefore, in this paper we address the question how the
thermodynamic properties of two leg spin-1/2 ladders are modified by
cyclic exchange interactions. In particular, the present work provides
the appropriate high temperature series (HTSE) data and results from
exact complete diagonalization (ED) \cite{fabri98a} of the magnetic
susceptibility and the specific heat. We expect that such an analysis
constitutes an important supplement to the study of spin-ladders,
furnishing additional information about couplings and interactions.

The model under study is given by the Hamiltonian
\begin{align}
  H & = \sum_{i}\Big(J_{\perp}\mathbf{S}_{1,i}\mathbf{S}_{2,i} +
  J_{\parallel}\big[\mathbf{S}_{1,i}\mathbf{S}_{1,i+1} +
  \mathbf{S}_{2,i}\mathbf{S}_{2,i+1}\big] \nonumber \\ +&\, 2J_{\text{cyc}}
  \big[(\mathbf{S}_{1,i}\mathbf{S}_{1,i+1})
  (\mathbf{S}_{2,i}\mathbf{S}_{2,i+1}) \label{eq:hamilton} \\ +&\,
  (\mathbf{S}_{1,i}\mathbf{S}_{2,i})
  (\mathbf{S}_{1,i+1}\mathbf{S}_{2,i+1}) -
  (\mathbf{S}_{1,i}\mathbf{S}_{2,i+1})
  (\mathbf{S}_{2,i}\mathbf{S}_{1,i+1})\big] \Big) \nonumber
\end{align}
where $J_{\perp}>0$ and $J_{\parallel}>0$ are the rung and leg
couplings, respectively; the subscript $i$ denotes the rungs and 1,2
the two legs. $J_{\text{cyc}}>0$ parametrizes the cyclic (4-spin) exchange.

In Sec.~\ref{sec:methods} we briefly sketch the methods we use. In
Sec.~\ref{sec:results}, results for the magnetic susceptibility
$\chi(T)$ 
\begin{equation}
  \label{eq:chi}
  \chi(\beta;J_{\parallel},J_{\text{cyc}}) =
  \frac{\beta}{N}\frac{\mathrm{tr}M^2e^{-\beta
      H}}{\mathrm{tr}e^{-\beta H}} = \frac{\beta}{N}\langle M^2
  \rangle
\end{equation}
and the specific heat $C(T)$ 
\begin{equation}
  \label{eq:heat}
  C(\beta;J_{\parallel},J_{\text{cyc}}) =
  \frac{1}{N}\frac{\partial}{\partial T} 
  \left( \frac{-\frac{\partial}{\partial \beta}\mathrm{tr}e^{-\beta
  H}}{ \mathrm{tr} e^{-\beta H}}\right)
\end{equation}
are presented for the Hamiltonian in Eq.~\ref{eq:hamilton}.
We also discuss representations of our results which can be
used to pinpoint the parameter regime from experimental data quickly.
\begin{figure}[htbp]
  \begin{center}
    \setlength{\unitlength}{1cm}
    \includegraphics[width=\columnwidth]{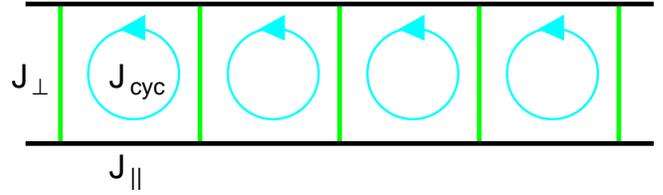}
    \caption{Two-leg ladder with cyclic (4-spin) exchange}
    \label{fig:ringladder}
  \end{center}
\end{figure}

\section{Methods\label{sec:methods}}
We use ED and the analytic method of HTSE. The computer
aided calculations for the HTSE yield polynomials in the coupling
parameters with fractions of integers as coefficients so that no
accuracy is lost. 

Details of our calculations can be found in Ref.~\cite{buehl00} and
details of the extrapolation schemes in Refs.~\cite{buehl01,bernu01}. Further,
the data are provided in electronic form so that it can be put to use quickly.

For both quantities $\chi(T)$ and $C(T)$ the results are provided up
to order ten in the dimensionless inverse temperature
$\beta=J_{\perp}/T$.  The first orders for $\chi(T)$ and $C(T)$ are
listed in Eqs. \ref{eq:chires} and \ref{eq:heatres}, where
$x_{\text{cyc}}=J_{\text{cyc}}/J_{\perp}$ and $x=J_{\parallel}/J_{\perp}$.
Higher orders are available electronically.

The bare truncated series are not sufficient to describe the
quantities under study at low values of $T$. Pad\'e representations
\cite{guttm89} are used to enhance decisively the region of validity of 
the HTSE results. Additionally, we include low temperature information
to enhance the region of validity even further. In some cases we can
describe the susceptibility in the complete temperature regime.  This
is in particular true if the system is substantially gapped because
the relevant correlation length remains finite restricted
by the inverse of $\text{max}(k_BT/\hbar v, \Delta/\hbar v)$, where
$v$ is a typical velocity of the excitations.

For a gapped system both $C(T)$ and $\chi(T)$ vanish for $T\rightarrow
0$. At finite but small temperatures the deviation from zero is
exponentially small due to the spin gap. Furthermore, the leading
power in $T$ depends only on the dimensionality of the system
\cite{troye94}. The low temperature susceptibility for the ladder
system can expressed by
\begin{subequations}
  \label{eq:lowT}
  \begin{equation}
    \label{eq:chilow}
    \chi(T) \approx {T}^{-\frac{1}{2}}e^{-\frac{\Delta}{T}}\ \text{
      for T}\ll \Delta \ .
  \end{equation}
  An analogous expression is obtained for $C(T)$
  \begin{equation}
    \label{eq:heatlow}
    C(T) \approx T^{-\frac{3}{2}}e^{-\frac{\Delta}{T}}\ \text{ for
      T}\ll \Delta \ .
  \end{equation}
\end{subequations}
Eqs.~\ref{eq:lowT} provide additional information to bias the
extrapolations. The value for the spin gap is obtained from
Ref.~\cite{schmipm}.

\begin{figure}[htbp]
  \setlength{\unitlength}{1cm}
  \begin{center}
    \leavevmode
    \includegraphics[angle=-90,width=\columnwidth]%
    {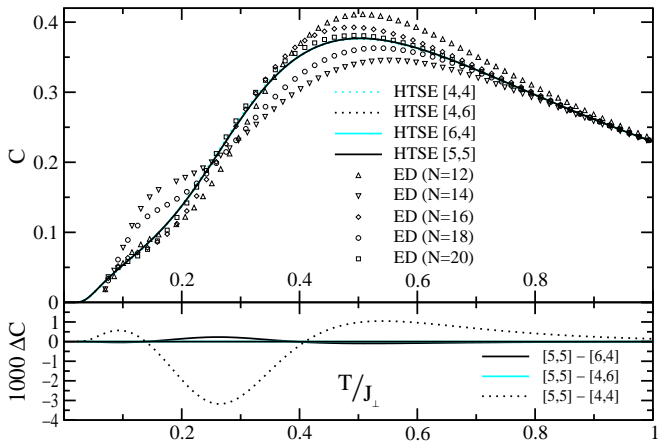}
    \caption{Specific heat for $x=1$ and
      $x_{\text{cyc}} = 0.1$, upper plot: $C(T)$ for various orders of
      Pad\'e representations and ED results from $N=12$ to $N=20$. lower
      plot: Differences between HTSE representations. The difference
      [5,5] - [4,6] cannot be discerned in the figure.} 
    \label{fig:spec_cmp}
  \end{center}
\end{figure}

In contrast to our previous extrapolation schemes \cite{buehl01}, here we
employ and extend the method suggested in Ref.~\cite{bernu01} to
extrapolate the HTSE data for the specific heat.  The main idea is to
express the entropy $S(T)$, obtained from the HTSE data of the
specific heat, in the new variable $e-e_0$, where $e_0$ is the ground
state energy and $e=e(T)$ is the average energy per site. The
temperature and the specific heat are then derived from the entropy as
functions of $e$. The ground state energy is obtained from
Ref.~\cite{schmipm}. To incorporate the low temperature information
from Eq.~\ref{eq:heatlow} the sum rule
\begin{equation}
  e-e_0 = \int_0^TC(T')dT' \approx T^{\frac{1}{2}}
  e^{-\frac{\Delta}{T}} 
  \label{eq:sumenergy}
\end{equation}
is considered in the limit $T\ll\Delta$. Inverting the above equation
provides an expression $T(e-e_0)$. Only an approximate solution is
possible. Taking the logarithm of Eq.~\ref{eq:sumenergy} and iterating
twice in $T$ leads to  
\begin{equation}
  T(y) \approx -\frac{\Delta}{\text{ln}(\sqrt{\Delta}y)}\ \text{ for
  }\ y\ll1 
  \label{eq:Tlow}
\end{equation}
with $y=e-e_0$. Combining Eqs.~\ref{eq:sumenergy} and \ref{eq:Tlow} and
the sum rule 
\begin{equation}
  S = \int_0^T\frac{C(T')}{T'}dT' \approx T^{-\frac{1}{2}}
  e^{-\frac{\Delta}{T}} \ \text{ for }\ T\ll\Delta
  \label{eq:sumentropy}
\end{equation}
yields the low temperature behavior of the entropy
\begin{equation}
  S(y) \approx -\frac{y}{\Delta}\text{ln}(\sqrt{\Delta}y))\ \text{ for
  } \ y\ll 1\ .
  \label{eq:entropylow}
\end{equation}
The logarithmic singularity at $e\!=\!e_0$ ($y\!=\!0$) can be avoided by
extrapolating the function
\begin{equation}
  \label{eq:Gapprox}
  G(y) = y\partial_y\frac{S(y)}{y}\ .
\end{equation}
The value of the gap $\Delta$ is incorporated by requiring
\begin{equation}
  \label{eq:Glow}
  G(y=0) = -1/\Delta \ .
\end{equation}
The entropy is then given by 
\begin{equation}
  \label{eq:entropyapprox}
  S(e) = (e-e_0)\Big(
  \int_0^e\frac{\tilde{G}(e')}{e'-e_0}de' -
  \frac{\text{ln}2}{e_0} \Big) 
\end{equation}
where $\tilde{G}$ is the Pad\'e approximant of $G$.  Wherever
possible diagonal Pad\'e representations, i.e. the same order in
numerator and denominator are used.  Exceptions to this rule will be
stated explicitly; they are necessary where spurious poles occur.
Comparison to ED results show that diagonal representations yield the
best results, see Fig.~\ref{fig:spec_cmp}. The convergence shown in
the lower plot is very convincing. The plot compares to ED data for
system sizes up to $N=20$. It is systematic to the ED calculations
that for increasing system sizes the results alternatingly yield an
upper ($N=12,\ 16,\ 20$) or a lower ($N=14,\ 18$) bound on the
specific heat for not too low temperatures. Therefore, the result
for $N=\infty$ should be in between the results with $N=18$ and
$N=20$, which is 
convincingly fulfilled by the HTSE extrapolations. The shoulders in
the low temperature regime at $T/J_{\perp}\lesssim 0.2$ are numerical
artifacts.

It has to be noted that the precise knowledge of the ground state
energy $e_0$ is of particular importance for the extrapolation of the
HTSE data for the specific heat. Even a small uncertainty of half a
percent in $e_0$ leads to significant differences in the specific heat
at and below its maximum. The high temperature part is unaffected
thereof.  The ground state energy $e_0(x,x_{\text{cyc}})$ is
calculated up to order $11$ in $x$ and $x_{\text{cyc}}$ in a high
order series expansion about the limit of isolated rungs. We set
$r=x_{\text{cyc}}/x=\ $const and use standard Dlog-Pad\'{e}
approximants on $de_0(x,rx)/dx$ which yield highly accurate results.

\begin{figure}[htbp]
  \setlength{\unitlength}{1cm}
  \begin{center}
    \leavevmode
    \includegraphics[angle=-90,width=\columnwidth]%
    {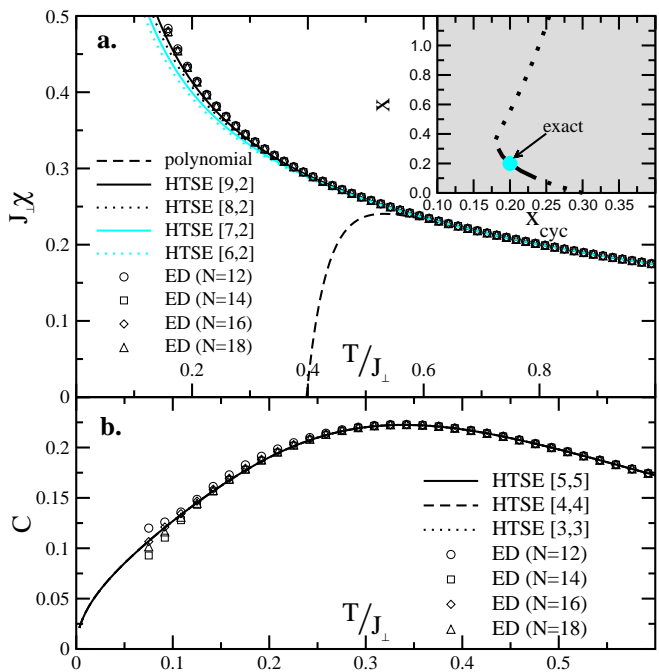}
    \caption{Susceptibility (a.) and specific heat (b.) for $x=0.2$
      and $x_{\text{cyc}}=0.2$ for various orders of Dlog-Pad\'e
      representations and ED results for $N=12,\ 14,\ 16$. HTSE
      representations in b. can not be resolved in the plot. Inset:
      Phase line with $\Delta=0$, for details see Ref.~\cite{schmipm}.
      }
    \label{fig:susc_cmp}
  \end{center}
\end{figure}
With the temperature as function of $e$ at hand it is possible to
represent also the susceptibility as function of $e$. The low
temperature behavior from Eq.~\ref{eq:chilow} can also be incorporated
in the extrapolations. Unfortunately, the convergence of the
extrapolations investigated is not as satisfying as for the
specific heat. The low temperature regime is underestimated and the
extrapolations are very sensitive to the order of numerator and
denominator in the Pad\'e representations. Most diagonal Pad\'e
approximants are not possible due to spurious poles. Hence we refrain
from using a $\chi(e)$ representation.

The extrapolation of the susceptibility follows Ref.~\cite{buehl01}.
Basically, the low temperature behavior from Eq.~\ref{eq:chilow} is
used to improve the representation. To incorporate the low temperature
information it is advantageous to map the temperature regime to the
interval $[0,1]$ via the substitution $u=\beta/(1+\beta)$. All
approximants are finite for $u\rightarrow 1$ and the representations
are no longer restricted to diagonal Dlog-Pad\'e approximants.  All
representations for $\chi(T)$ are extrapolated with the same order of
Dlog-Pad\'e approximants to retain a consistent description of the
HTSE results.  We use the $[n,2]$-Dlog-Pad\'e representation in $u$
with the only restriction $n+2\le11$. The representation chosen is
checked in the limit of the (isotropic) ladder and the Heisenberg
chain, where precise \cite{oitma96,weiho97,frisc96} or exact
\cite{klump93b} results are available.

To assess the range of validity we investigate various orders of
Dlog-Pad\'e approximants. They are compared to the highest order
available. The convergence is very satisfying. The orders 8 to 10
differ only by $10^{-3}T/J_{\perp}$ from the 11th order for $\chi$
(not shown).  This observation does not change significantly for other
sets of parameters considered.

Another important check is the comparison to ED results.  In
Fig.~\ref{fig:susc_cmp} data for the susceptibility and the specific
heat are shown for the
gapless point $x = 0.2$ and $x_{\text{cyc}}=0.2$, which is exactly
known \cite{schmipm}. The inset shows the phase line where the gap
vanishes. The results are obtained by a high order series expansion
about the limit of isolated rungs. The solid line about the exact
point shows the highly convergent results. The dotted lines give a
sketch of the phase line for parameters far away from the exact
point. The phase diagram is investigated in detail in
Ref.~\cite{schmipm}. Left to the phase line the system is in a gapped
rung singlet phase. On increasing $x_{\text{cyc}}$ the gap vanishes
linearly in $x_{\text{cyc}}$ and opens again linearly on the right of
the phase line in a staggered dimer phase \cite{muell02,laeuc02}.

To derive the low temperature behavior of $C$ and $\chi$ at the
gapless point the exactly known triplet dispersion $\omega (q) \propto
q^2$ at $q=\pi$ \cite{schmipm} is used. A similar analysis as in
Ref.~\cite{troye94} yields
\begin{subequations}
  \label{eq:Tlow_gap0}
  \begin{eqnarray}
    \chi(T) &\approx& \frac{1}{\sqrt{T}}\ \text{ for } T\ll1 
    \label{eq:chilow_gap0} \\
    C(T) &\approx& \sqrt{T}\ \text{ for } T\ll1\ . 
    \label{eq:Clow_gap0}
  \end{eqnarray}
\end{subequations}
Using the sum rules mentioned above for the specific heat leads to an
entropy 
\begin{equation}
  \label{eq:Slow_gap0}
  S(e) \approx (e-e_0)^{1/3}\ \text{ for }\ e-e_0\ \text{ small.}
\end{equation}
The entropy is extrapolated using a Dlog-Pad\'e approximant biased to
contain the extra information from Eq.~\ref{eq:Slow_gap0}. 

The chosen representations of the HTSE results yield stable and
trustworthy results for the calculated thermodynamical properties.
Especially the experimentally interesting position and height of the
maximum are sufficiently well described for quantitative predictions.

\begin{widetext}
  \begin{subequations}
    \begin{eqnarray}
      4T\chi 
      &=& 1+\left(-\textstyle{\frac{1}{4}}
        -\textstyle{\frac{1}{2}}x\right)\beta +
      \left(\left(\textstyle{\frac{1}{4}}
          +\textstyle{\frac{3}{16}}x_{\text{cyc}}\right)x
        -\textstyle{\frac{7}{32}}x_{\text{cyc}}^2 -\textstyle{\frac{1}{16}}
        +\textstyle{\frac{3}{16}}x_{\text{cyc}}\right)\beta^2
      +\mathcal{O}(\beta^3) \label{eq:chires} \\ 
      16C 
      &=& \left( \textstyle{\frac{3}{2}} +3x^2
        +\textstyle{\frac{21}{8}}x_{\text{cyc}}^2\right)\beta^2 +
      \left(\textstyle{\frac{3}{4}}
        +\textstyle{\frac{3}{2}}x^3
        -\textstyle{\frac{27}{8}}x_{\text{cyc}}
        -\textstyle{\frac{27}{8}}x_{\text{cyc}}x^2
        +\textstyle{\frac{45}{8}}x_{\text{cyc}}^2
        +\textstyle{\frac{45}{8}}x_{\text{cyc}}^2x
        -\textstyle{\frac{9}{8}}x_{\text{cyc}}^3\right)\beta^3
      +\mathcal{O}(\beta^4)
      \label{eq:heatres}
    \end{eqnarray}
  \end{subequations}
\end{widetext}
\section{Results\label{sec:results}}
Our aim is to provide results which show the quantitative behavior of
the thermodynamic properties and in particular the effects
of a cyclic spin exchange. With the help of computer algebra programs
the HTSE results can be used easily to determine the model parameters
of a substance under consideration. Only data of standard quantities
like the magnetic susceptibility is necessary. The occurring ambiguity
in determining the model parameters (\cite{buehl01} and see below) from
only one quantity like $\chi(T)$ can be resolved by the knowledge of
other quantities, e.g. the spin gap $\Delta$ or the magnetic specific
heat $C(T)$ as far as accessible experimentally.
\begin{figure}[htbp]
  \begin{center}
    \leavevmode
    \includegraphics[angle=-90,width=\columnwidth]%
    {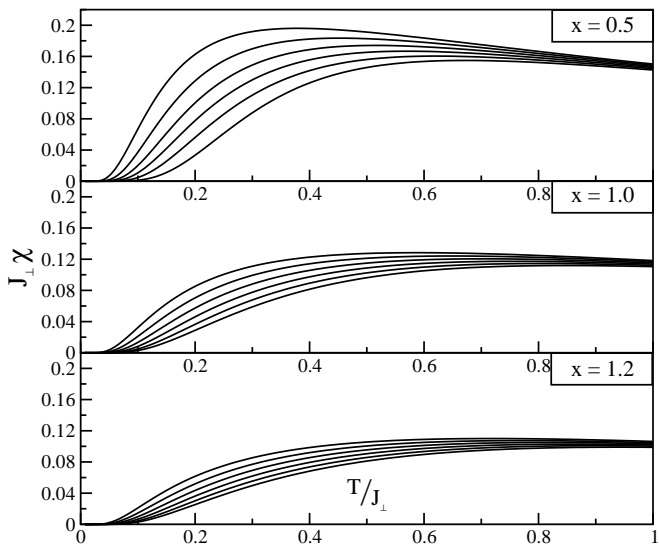}
    \caption{$\chi(T)$ for various values of $x$
      for $x_{\text{cyc}}=0,0.02,0.04,0.06,0.08$ and $0.1$ in ascending
      order from right to left.  }
    \label{fig:susc_all}
  \end{center}
\end{figure}

\subsection{\label{sec:results:chi} Susceptibility}
Fig.~\ref{fig:susc_all} shows an overview of the magnetic
susceptibility for various values of the cyclic exchange $x_{\text{cyc}}$ and the
leg coupling $x$. The choice of the shown parameter regime
is taken from published values for substances presently investigated
\cite{matsu00b,schmi01,nunne02}.

A general behavior for increasing $J_{\text{cyc}}$ at fixed $J_{\parallel}$ is
the shift of an increasing $\chi_{\text{max}}$ to lower temperatures.
This effect is induced by the decrease of the whole dispersion, i.e.
all energies are lowered \cite{brehm99,schmipm} and the global $1/T$
factor which enhances the value of $\chi_{\text{max}}$. For increasing
$x$ this effect is weakened. The increasing leg coupling
provides an additional antiferromagnetic coupling stabilizing the
system against magnetic perturbations. Thus, an increasing
$x$ stabilizes the system and counteracts an increasing
$x_{\text{cyc}}$ which destabilizes it.
\begin{figure}[htbp]
  \begin{center}
    \leavevmode
    \includegraphics[angle=-90,width=\columnwidth]{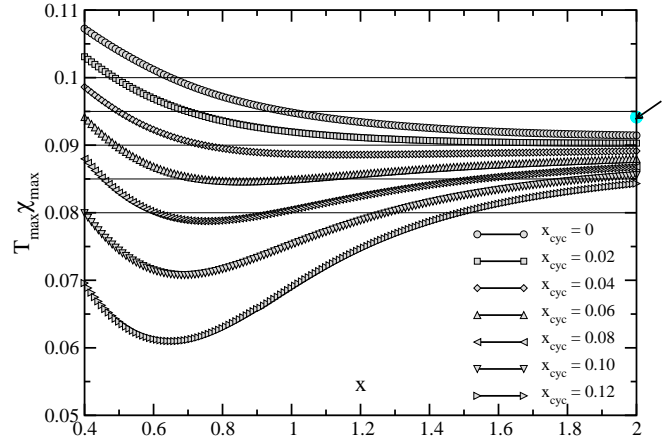}
    \caption{$\chi_{\text{max}}T_{\text{max}}$ versus $x$
    for $x_{\text{cyc}}=0,0.02,0.04,0.06,0.08$ and $0.1$ in descending
    order. The symbols are the calculated data points from the
    HTSE. The filled circle on the right side indicates the
    $\chi_{\text{max}}T_{\text{max}}$ value for an isotropic chain which
    is known exactly \cite{klump93b}. Horizontal lines show the
    constant values used to rescale $\chi(T)$ in
    Fig.~\ref{fig:scaled_chi}.}  
    \label{fig:TmaxChimax}
  \end{center}
\end{figure}

\begin{figure}[htbp]
  \begin{center}
    \leavevmode
    \includegraphics[angle=-90,width=\columnwidth]%
     {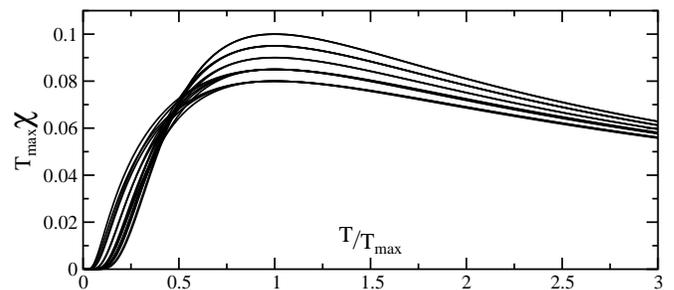}
    \caption{Rescaled susceptibilities for
    $\chi_{\text{max}}T_{\text{max}}=0.1,\ 0.095,\ 0.09,\ 0.085,\
    0.08$ for the $x$ and $x_{\text{cyc}}$ values shown in
    Fig. \ref{fig:TmaxChimax}}
    \label{fig:scaled_chi}
  \end{center}
\end{figure}

In Figs.~\ref{fig:TmaxChimax} and \ref{fig:scaled_chi} we address the
information content of a measurement of $\chi(T)$
(cf. \cite{buehl01}). Fig.~\ref{fig:TmaxChimax} shows the energy-scale
independent quantity $\chi_{\text{max}}T_{\text{max}}$ which is a
characteristic in experimental measurements.  Note that for increasing
$x$ the differences between the curves for various values of
$x_{\text{cyc}}$ become smaller, because $J_{\parallel}$ and
$J_{\perp}$ set the changing energy scale, whereas $J_{\text{cyc}}$
stays constant. In the limit of large $x$ the system approaches two
independent chains with a decreasing relative interchain coupling
induced by $J_{\perp}$ and $J_{\text{cyc}}$.

Once the value of $\chi_{\text{max}}T_{\text{max}}$ is measured the
parameter set ($x$,$x_{\text{cyc}}$) can be read off from
Fig.~\ref{fig:TmaxChimax}. But there is still an ambiguity which
cannot be resolved by a measurement of $\chi(T)$ alone as illustrated
in Fig.~\ref{fig:scaled_chi}. 
There the rescaled susceptibilities are depicted for the indicated values
of $\chi_{\text{max}}T_{\text{max}}$ in Fig.~\ref{fig:TmaxChimax}
(solid horizontal lines). The main feature in the susceptibility
curves is the maximum. For different sets of parameters for a specific
value of $\chi_{\text{max}}T_{\text{max}}$ the qualitative and
quantitative behavior cannot be distinguished unless precise
measurements in the low temperature regime are possible.  

To summarize the latter investigations we state that with the present
results of ED and HTSE it is difficult to ascertain all model
parameters from the temperature dependence of the susceptibility
alone, see also discussion in Ref.~\cite{buehl01}. With the knowledge
of detailed low temperature information this problem can be resolved.

\begin{figure}[htbp]
  \setlength{\unitlength}{1cm}
  \begin{center}
    \leavevmode
    \includegraphics[angle=-90,width=\columnwidth]%
    {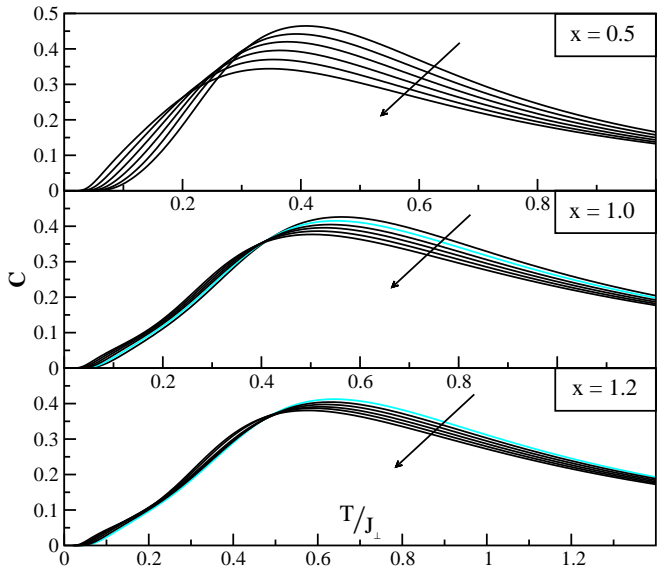}
    \caption{$C(T)$ for three values for $x$ with
      $x_{\text{cyc}}=0,0.02,0.04,0.06,0.08$ and $0.1$ in ascending
      order in direction of arrows. Gray lines show [6,4] Pad\'e
      representations. 
      }
    \label{fig:call}
  \end{center}
\end{figure}
\subsection{\label{sec:results:heat} Specific heat}
Further information about the magnetic properties of 
substances can be obtained by measuring also the magnetic specific heat
$C(T)$. We are aware, however, that it is difficult to extract the
magnetic contribution from the measured specific 
heat if the energy scale of the lattice vibrations is of the same
order as the magnetic couplings.

Once the magnetic specific heat is known the ambiguity in 
parameters can be resolved. Fig.~\ref{fig:call} shows an overview
of the magnetic specific heat for $x=0.5,\ 1,\
1.2$ and $x_{\text{cyc}}=0,\ldots,0.1$. For increasing leg
coupling $x$ and fixed $x_{\text{cyc}}$ the position of
$C_{\text{max}}$ shifts to higher temperatures and the height lowers
slightly for $x_{\text{cyc}}=0,0.02$, it stays almost constant for
$x_{\text{cyc}}=0.04,0.06$, and it increases slightly for
$x_{\text{cyc}}=0.08,0.1$. For increasing cyclic exchange $x_{\text{cyc}}$ and
fixed leg coupling $C_{\text{max}}$ moves to lower temperatures and
decreases. This behavior is induced by the decreasing overall
dispersion, as discussed above.

\section{Conclusions\label{sec:conclusions}}
We investigated the thermodynamic   properties of the
two-leg spin-1/2 ladder with cyclic exchange. We used two methods, ED
and HTSE. The representation of the HTSE results were optimized by
using Dlog-Pad\'e and Pad\'e approximants including the behavior of the
considered quantities, $\chi(T)$ and $C(T)$ at low temperatures.

The results can serve as input for quick and easy data analysis to
determine the model parameters. Especially the experimentally
interesting position and height of the maxima of $\chi(T)$ and $C(T)$
are described quantitatively.
 
We showed that with the measurement of one quantity alone, e.g. the
magnetic susceptibility it is difficult to determine all model
parameters. Additional information like the spin gap or the specific
heat helps fixing the parameters.

\begin{acknowledgments}
The computations were done in the Regionales Rechenzentrum der
Universit\"at zu K\"oln with the kind support of P.~Br\"uhne. This work
was supported by the DFG by Schwerpunkt 1073 and by SFB 608.
\end{acknowledgments}


\begin{thebibliography}{30}
\expandafter\ifx\csname natexlab\endcsname\relax\def\natexlab#1{#1}\fi
\expandafter\ifx\csname bibnamefont\endcsname\relax
  \def\bibnamefont#1{#1}\fi
\expandafter\ifx\csname bibfnamefont\endcsname\relax
  \def\bibfnamefont#1{#1}\fi
\expandafter\ifx\csname citenamefont\endcsname\relax
  \def\citenamefont#1{#1}\fi
\expandafter\ifx\csname url\endcsname\relax
  \def\url#1{\texttt{#1}}\fi
\expandafter\ifx\csname urlprefix\endcsname\relax\def\urlprefix{URL }\fi
\providecommand{\bibinfo}[2]{#2}
\providecommand{\eprint}[2][]{\url{#2}}

\bibitem[{\citenamefont{Dagotto and Rice}(1996)}]{dagot96}
\bibinfo{author}{\bibfnamefont{E.}~\bibnamefont{Dagotto}} \bibnamefont{and}
  \bibinfo{author}{\bibfnamefont{T.~M.} \bibnamefont{Rice}},
  \bibinfo{journal}{Science} \textbf{\bibinfo{volume}{271}},
  \bibinfo{pages}{618} (\bibinfo{year}{1996}).

\bibitem[{\citenamefont{Schmidt and Kuramoto}(1990)}]{schmi90}
\bibinfo{author}{\bibfnamefont{H.~J.} \bibnamefont{Schmidt}} \bibnamefont{and}
  \bibinfo{author}{\bibfnamefont{Y.}~\bibnamefont{Kuramoto}},
  \bibinfo{journal}{Physica C} \textbf{\bibinfo{volume}{167}},
  \bibinfo{pages}{263} (\bibinfo{year}{1990}).

\bibitem[{\citenamefont{Honda et~al.}(1993)\citenamefont{Honda, Kuramoto, and
  Watanabe}}]{honda93}
\bibinfo{author}{\bibfnamefont{Y.}~\bibnamefont{Honda}},
  \bibinfo{author}{\bibfnamefont{Y.}~\bibnamefont{Kuramoto}}, \bibnamefont{and}
  \bibinfo{author}{\bibfnamefont{T.}~\bibnamefont{Watanabe}},
  \bibinfo{journal}{Phys. Rev. B} \textbf{\bibinfo{volume}{47}},
  \bibinfo{pages}{11329} (\bibinfo{year}{1993}).

\bibitem[{\citenamefont{Brehmer et~al.}(1999)\citenamefont{Brehmer, Mikeska,
  M{\"u}ller, Nagaosa, and Uchida}}]{brehm99}
\bibinfo{author}{\bibfnamefont{S.}~\bibnamefont{Brehmer}},
  \bibinfo{author}{\bibfnamefont{H.-J.} \bibnamefont{Mikeska}},
  \bibinfo{author}{\bibfnamefont{M.}~\bibnamefont{M{\"u}ller}},
  \bibinfo{author}{\bibfnamefont{N.}~\bibnamefont{Nagaosa}}, \bibnamefont{and}
  \bibinfo{author}{\bibfnamefont{S.}~\bibnamefont{Uchida}},
  \bibinfo{journal}{Phys. Rev. B} \textbf{\bibinfo{volume}{60}},
  \bibinfo{pages}{329} (\bibinfo{year}{1999}).

\bibitem[{\citenamefont{Reischl and M\"uller-Hartmann}(2002)}]{reisc01}
\bibinfo{author}{\bibfnamefont{A.}~\bibnamefont{Reischl}} \bibnamefont{and}
  \bibinfo{author}{\bibfnamefont{E.}~\bibnamefont{M\"uller-Hartmann}},
  \bibinfo{journal}{Eur. Phys. J. B} \textbf{\bibinfo{volume}{28}},
  \bibinfo{pages}{173} (\bibinfo{year}{2002}).

\bibitem[{\citenamefont{Senthil and Fisher}(2001)}]{senth01a}
\bibinfo{author}{\bibfnamefont{T.}~\bibnamefont{Senthil}} \bibnamefont{and}
  \bibinfo{author}{\bibfnamefont{M.~P.~A.} \bibnamefont{Fisher}},
  \bibinfo{journal}{Phys. Rev. Lett.} \textbf{\bibinfo{volume}{86}},
  \bibinfo{pages}{292} (\bibinfo{year}{2001}).

\bibitem[{\citenamefont{Balents et~al.}(2002)\citenamefont{Balents, Fisher, and
  Girvin}}]{balen02}
\bibinfo{author}{\bibfnamefont{L.}~\bibnamefont{Balents}},
  \bibinfo{author}{\bibfnamefont{M.~P.~A.} \bibnamefont{Fisher}},
  \bibnamefont{and} \bibinfo{author}{\bibfnamefont{S.~M.}
  \bibnamefont{Girvin}}, \bibinfo{journal}{Phys. Rev. B}
  \textbf{\bibinfo{volume}{65}}, \bibinfo{pages}{224412}
  (\bibinfo{year}{2002}).

\bibitem[{\citenamefont{Lorenzana et~al.}(1999)\citenamefont{Lorenzana, Eroles,
  and Sorella}}]{loren99}
\bibinfo{author}{\bibfnamefont{J.}~\bibnamefont{Lorenzana}},
  \bibinfo{author}{\bibfnamefont{J.}~\bibnamefont{Eroles}}, \bibnamefont{and}
  \bibinfo{author}{\bibfnamefont{S.}~\bibnamefont{Sorella}},
  \bibinfo{journal}{Phys. Rev. Lett.} \textbf{\bibinfo{volume}{83}},
  \bibinfo{pages}{5122} (\bibinfo{year}{1999}).

\bibitem[{\citenamefont{Mizuno et~al.}(1999)\citenamefont{Mizuno, Tohyama, and
  Maekawa}}]{mizun99}
\bibinfo{author}{\bibfnamefont{Y.}~\bibnamefont{Mizuno}},
  \bibinfo{author}{\bibfnamefont{T.}~\bibnamefont{Tohyama}}, \bibnamefont{and}
  \bibinfo{author}{\bibfnamefont{S.}~\bibnamefont{Maekawa}},
  \bibinfo{journal}{J. Low Temp. Phys.} \textbf{\bibinfo{volume}{117}},
  \bibinfo{pages}{389} (\bibinfo{year}{1999}).

\bibitem[{\citenamefont{Matsuda et~al.}(2000)\citenamefont{Matsuda, Katsumata,
  Eccleston, Brehmer, and Mikeska}}]{matsu00b}
\bibinfo{author}{\bibfnamefont{M.}~\bibnamefont{Matsuda}},
  \bibinfo{author}{\bibfnamefont{K.}~\bibnamefont{Katsumata}},
  \bibinfo{author}{\bibfnamefont{R.~S.} \bibnamefont{Eccleston}},
  \bibinfo{author}{\bibfnamefont{S.}~\bibnamefont{Brehmer}}, \bibnamefont{and}
  \bibinfo{author}{\bibfnamefont{H.-J.} \bibnamefont{Mikeska}},
  \bibinfo{journal}{Phys. Rev. B} \textbf{\bibinfo{volume}{62}},
  \bibinfo{pages}{8903} (\bibinfo{year}{2000}).

\bibitem[{\citenamefont{LiMing et~al.}(2000)\citenamefont{LiMing, Misguich,
  Sindzingre, and Lhuillier}}]{limin00}
\bibinfo{author}{\bibfnamefont{W.}~\bibnamefont{LiMing}},
  \bibinfo{author}{\bibfnamefont{G.}~\bibnamefont{Misguich}},
  \bibinfo{author}{\bibfnamefont{P.}~\bibnamefont{Sindzingre}},
  \bibnamefont{and}
  \bibinfo{author}{\bibfnamefont{C.}~\bibnamefont{Lhuillier}},
  \bibinfo{journal}{Phys. Rev. B} \textbf{\bibinfo{volume}{62}},
  \bibinfo{pages}{6372} (\bibinfo{year}{2000}).

\bibitem[{\citenamefont{Schmidt et~al.}(2001)\citenamefont{Schmidt, Knetter,
  and Uhrig}}]{schmi01}
\bibinfo{author}{\bibfnamefont{K.~P.} \bibnamefont{Schmidt}},
  \bibinfo{author}{\bibfnamefont{C.}~\bibnamefont{Knetter}}, \bibnamefont{and}
  \bibinfo{author}{\bibfnamefont{G.~S.} \bibnamefont{Uhrig}},
  \bibinfo{journal}{Europhys. Lett.} \textbf{\bibinfo{volume}{56}},
  \bibinfo{pages}{877} (\bibinfo{year}{2001}).

\bibitem[{\citenamefont{Katanin and Kampf}(2002)}]{katan01}
\bibinfo{author}{\bibfnamefont{A.~A.} \bibnamefont{Katanin}} \bibnamefont{and}
  \bibinfo{author}{\bibfnamefont{A.~P.} \bibnamefont{Kampf}},
  \bibinfo{journal}{Phys. Rev. B} \textbf{\bibinfo{volume}{66}},
  \bibinfo{pages}{100403} (\bibinfo{year}{2002}).

\bibitem[{\citenamefont{Coldea et~al.}(2001)\citenamefont{Coldea, Hayden,
  Aeppli, Perring, Frost, Mason, Cheong, and Fisk}}]{colde01}
\bibinfo{author}{\bibfnamefont{R.}~\bibnamefont{Coldea}},
  \bibinfo{author}{\bibfnamefont{S.~M.} \bibnamefont{Hayden}},
  \bibinfo{author}{\bibfnamefont{G.}~\bibnamefont{Aeppli}},
  \bibinfo{author}{\bibfnamefont{T.~G.} \bibnamefont{Perring}},
  \bibinfo{author}{\bibfnamefont{C.~D.} \bibnamefont{Frost}},
  \bibinfo{author}{\bibfnamefont{T.~E.} \bibnamefont{Mason}},
  \bibinfo{author}{\bibfnamefont{S.-W.} \bibnamefont{Cheong}},
  \bibnamefont{and} \bibinfo{author}{\bibfnamefont{Z.}~\bibnamefont{Fisk}},
  \bibinfo{journal}{Phys. Rev. Lett.} \textbf{\bibinfo{volume}{86}},
  \bibinfo{pages}{5377} (\bibinfo{year}{2001}).

\bibitem[{\citenamefont{Windt et~al.}(2001)\citenamefont{Windt, Gr\"uninger,
  Nunner, Knetter, Schmidt, Uhrig, Kopp, Freimuth, Ammerahl, B{\"u}chner
  et~al.}}]{windt01}
\bibinfo{author}{\bibfnamefont{M.}~\bibnamefont{Windt}},
  \bibinfo{author}{\bibfnamefont{M.}~\bibnamefont{Gr\"uninger}},
  \bibinfo{author}{\bibfnamefont{T.}~\bibnamefont{Nunner}},
  \bibinfo{author}{\bibfnamefont{C.}~\bibnamefont{Knetter}},
  \bibinfo{author}{\bibfnamefont{K.}~\bibnamefont{Schmidt}},
  \bibinfo{author}{\bibfnamefont{G.~S.} \bibnamefont{Uhrig}},
  \bibinfo{author}{\bibfnamefont{T.}~\bibnamefont{Kopp}},
  \bibinfo{author}{\bibfnamefont{A.}~\bibnamefont{Freimuth}},
  \bibinfo{author}{\bibfnamefont{U.}~\bibnamefont{Ammerahl}},
  \bibinfo{author}{\bibfnamefont{B.}~\bibnamefont{B{\"u}chner}},
  \bibnamefont{et~al.}, \bibinfo{journal}{Phys. Rev. Lett.}
  \textbf{\bibinfo{volume}{87}}, \bibinfo{pages}{127002}
  (\bibinfo{year}{2001}).

\bibitem[{\citenamefont{Nunner et~al.}(in press)\citenamefont{Nunner, Brune,
  Kopp, Windt, and Gr{\"u}ninger}}]{nunne02}
\bibinfo{author}{\bibfnamefont{T.~S.} \bibnamefont{Nunner}},
  \bibinfo{author}{\bibfnamefont{P.}~\bibnamefont{Brune}},
  \bibinfo{author}{\bibfnamefont{T.}~\bibnamefont{Kopp}},
  \bibinfo{author}{\bibfnamefont{M.}~\bibnamefont{Windt}}, \bibnamefont{and}
  \bibinfo{author}{\bibfnamefont{M.}~\bibnamefont{Gr{\"u}ninger}},
  \bibinfo{journal}{Phys. Rev. B, Rapid Comm.}  (\bibinfo{year}{in press}).

\bibitem[{\citenamefont{Johnston et~al.}(2000)\citenamefont{Johnston, Troyer,
  Miyahara, Lidsky, Ueda, Azuma, Hiroi, Takano, Isobe, Ueda et~al.}}]{johns00a}
\bibinfo{author}{\bibfnamefont{D.~C.} \bibnamefont{Johnston}},
  \bibinfo{author}{\bibfnamefont{M.}~\bibnamefont{Troyer}},
  \bibinfo{author}{\bibfnamefont{S.}~\bibnamefont{Miyahara}},
  \bibinfo{author}{\bibfnamefont{D.}~\bibnamefont{Lidsky}},
  \bibinfo{author}{\bibfnamefont{K.}~\bibnamefont{Ueda}},
  \bibinfo{author}{\bibfnamefont{M.}~\bibnamefont{Azuma}},
  \bibinfo{author}{\bibfnamefont{Z.}~\bibnamefont{Hiroi}},
  \bibinfo{author}{\bibfnamefont{M.}~\bibnamefont{Takano}},
  \bibinfo{author}{\bibfnamefont{M.}~\bibnamefont{Isobe}},
  \bibinfo{author}{\bibfnamefont{Y.}~\bibnamefont{Ueda}}, \bibnamefont{et~al.},
  \bibinfo{journal}{cond-mat/0001147}  (\bibinfo{year}{2000}).

\bibitem[{\citenamefont{Fabricius et~al.}(1998)\citenamefont{Fabricius,
  Kl\"umper, L\"ow, B\"uchner, Lorenz, Dhalenne, and Revcolevschi}}]{fabri98a}
\bibinfo{author}{\bibfnamefont{K.}~\bibnamefont{Fabricius}},
  \bibinfo{author}{\bibfnamefont{A.}~\bibnamefont{Kl\"umper}},
  \bibinfo{author}{\bibfnamefont{U.}~\bibnamefont{L\"ow}},
  \bibinfo{author}{\bibfnamefont{B.}~\bibnamefont{B\"uchner}},
  \bibinfo{author}{\bibfnamefont{T.}~\bibnamefont{Lorenz}},
  \bibinfo{author}{\bibfnamefont{G.}~\bibnamefont{Dhalenne}}, \bibnamefont{and}
  \bibinfo{author}{\bibfnamefont{A.}~\bibnamefont{Revcolevschi}},
  \bibinfo{journal}{Phys. Rev. B} \textbf{\bibinfo{volume}{57}},
  \bibinfo{pages}{1102} (\bibinfo{year}{1998}).

\bibitem[{\citenamefont{B\"uhler et~al.}(2000)\citenamefont{B\"uhler, Elstner,
  and Uhrig}}]{buehl00}
\bibinfo{author}{\bibfnamefont{A.}~\bibnamefont{B\"uhler}},
  \bibinfo{author}{\bibfnamefont{N.}~\bibnamefont{Elstner}}, \bibnamefont{and}
  \bibinfo{author}{\bibfnamefont{G.~S.} \bibnamefont{Uhrig}},
  \bibinfo{journal}{Eur. Phys. J. B} \textbf{\bibinfo{volume}{16}},
  \bibinfo{pages}{475} (\bibinfo{year}{2000}).

\bibitem[{\citenamefont{B\"uhler et~al.}(2001)\citenamefont{B\"uhler, L\"ow,
  and Uhrig}}]{buehl01}
\bibinfo{author}{\bibfnamefont{A.}~\bibnamefont{B\"uhler}},
  \bibinfo{author}{\bibfnamefont{U.}~\bibnamefont{L\"ow}}, \bibnamefont{and}
  \bibinfo{author}{\bibfnamefont{G.~S.} \bibnamefont{Uhrig}},
  \bibinfo{journal}{Phys. Rev. B} \textbf{\bibinfo{volume}{64}},
  \bibinfo{pages}{024428} (\bibinfo{year}{2001}).

\bibitem[{\citenamefont{Bernu and Misguich}(2001)}]{bernu01}
\bibinfo{author}{\bibfnamefont{B.}~\bibnamefont{Bernu}} \bibnamefont{and}
  \bibinfo{author}{\bibfnamefont{G.}~\bibnamefont{Misguich}},
  \bibinfo{journal}{Phys. Rev. B} \textbf{\bibinfo{volume}{63}},
  \bibinfo{pages}{134409} (\bibinfo{year}{2001}).

\bibitem[{\citenamefont{Guttmann}(1989)}]{guttm89}
\bibinfo{author}{\bibfnamefont{A.}~\bibnamefont{Guttmann}},
  \emph{\bibinfo{title}{Phase Transition and Critical Phenomena}}
  (\bibinfo{publisher}{Academic Press}, \bibinfo{address}{New York},
  \bibinfo{year}{1989}), vol.~\bibinfo{volume}{13}, chap.~\bibinfo{chapter}{1}.

\bibitem[{\citenamefont{Troyer et~al.}(1994)\citenamefont{Troyer, Tsunetsugu,
  and W\"urtz}}]{troye94}
\bibinfo{author}{\bibfnamefont{M.}~\bibnamefont{Troyer}},
  \bibinfo{author}{\bibfnamefont{H.}~\bibnamefont{Tsunetsugu}},
  \bibnamefont{and} \bibinfo{author}{\bibfnamefont{D.}~\bibnamefont{W\"urtz}},
  \bibinfo{journal}{Phys. Rev. B} \textbf{\bibinfo{volume}{50}},
  \bibinfo{pages}{13515} (\bibinfo{year}{1994}), \bibinfo{note}{{T}here is a
  misprint in Eqs. (38) and (39), $T/\Delta$ must be replaced by $\Delta/T$}.

\bibitem[{\citenamefont{Schmidt et~al.}()\citenamefont{Schmidt, Monien, and
  Uhrig}}]{schmipm}
\bibinfo{author}{\bibfnamefont{K.~P.} \bibnamefont{Schmidt}},
  \bibinfo{author}{\bibfnamefont{H.}~\bibnamefont{Monien}}, \bibnamefont{and}
  \bibinfo{author}{\bibfnamefont{G.~S.} \bibnamefont{Uhrig}}, \bibinfo{note}{to
  be published}.

\bibitem[{\citenamefont{Oitmaa et~al.}(1996)\citenamefont{Oitmaa, Singh, and
  Weihong}}]{oitma96}
\bibinfo{author}{\bibfnamefont{J.}~\bibnamefont{Oitmaa}},
  \bibinfo{author}{\bibfnamefont{R.~R.~P.} \bibnamefont{Singh}},
  \bibnamefont{and} \bibinfo{author}{\bibfnamefont{Z.}~\bibnamefont{Weihong}},
  \bibinfo{journal}{Phys. Rev. B} \textbf{\bibinfo{volume}{54}},
  \bibinfo{pages}{1009} (\bibinfo{year}{1996}).

\bibitem[{\citenamefont{Weihong et~al.}(1997)\citenamefont{Weihong, Singh, and
  Oitmaa}}]{weiho97}
\bibinfo{author}{\bibfnamefont{Z.}~\bibnamefont{Weihong}},
  \bibinfo{author}{\bibfnamefont{R.~R.~P.} \bibnamefont{Singh}},
  \bibnamefont{and} \bibinfo{author}{\bibfnamefont{J.}~\bibnamefont{Oitmaa}},
  \bibinfo{journal}{Phys. Rev. B} \textbf{\bibinfo{volume}{55}},
  \bibinfo{pages}{8052} (\bibinfo{year}{1997}).

\bibitem[{\citenamefont{Frischmuth et~al.}(1996)\citenamefont{Frischmuth,
  Ammon, and Troyer}}]{frisc96}
\bibinfo{author}{\bibfnamefont{B.}~\bibnamefont{Frischmuth}},
  \bibinfo{author}{\bibfnamefont{B.}~\bibnamefont{Ammon}}, \bibnamefont{and}
  \bibinfo{author}{\bibfnamefont{M.}~\bibnamefont{Troyer}},
  \bibinfo{journal}{Phys. Rev. B} \textbf{\bibinfo{volume}{54}},
  \bibinfo{pages}{R3714} (\bibinfo{year}{1996}).

\bibitem[{\citenamefont{Kl\"umper}(1993)}]{klump93b}
\bibinfo{author}{\bibfnamefont{A.}~\bibnamefont{Kl\"umper}},
  \bibinfo{journal}{Z. Phys. B} \textbf{\bibinfo{volume}{91}},
  \bibinfo{pages}{507} (\bibinfo{year}{1993}).

\bibitem[{\citenamefont{M{\"u}ller et~al.}(2002)\citenamefont{M{\"u}ller,
  Vekua, and Mikeska}}]{muell02}
\bibinfo{author}{\bibfnamefont{M.}~\bibnamefont{M{\"u}ller}},
  \bibinfo{author}{\bibfnamefont{T.}~\bibnamefont{Vekua}}, \bibnamefont{and}
  \bibinfo{author}{\bibfnamefont{H.-J.} \bibnamefont{Mikeska}},
  \bibinfo{journal}{cond-mat/0206081}  (\bibinfo{year}{2002}).

\bibitem[{\citenamefont{L{\"a}uchli et~al.}(2002)\citenamefont{L{\"a}uchli,
  Schmid, and Troyer}}]{laeuc02}
\bibinfo{author}{\bibfnamefont{A.}~\bibnamefont{L{\"a}uchli}},
  \bibinfo{author}{\bibfnamefont{G.}~\bibnamefont{Schmid}}, \bibnamefont{and}
  \bibinfo{author}{\bibfnamefont{M.}~\bibnamefont{Troyer}},
  \bibinfo{journal}{cond-mat/0206153}  (\bibinfo{year}{2002}).

\end{thebibliography}
\end{document}